\documentclass[11pt, a4paper, logo, copyright]{deepmind}

\usepackage[authoryear, sort&compress, round]{natbib}
\usepackage{dsfont}        
\usepackage{marginnote}   

\usepackage{exscale}       
\newcommand{\norm}[1]{\left\lVert#1\right\rVert}
\newcommand{\R}[0]{\mathds{R}}

\renewcommand{\vec}[1]{\mathbf{#1}}

\title{Inferring a Continuous Distribution of Atom Coordinates from Cryo-EM Images using VAEs}

\correspondingauthor{jonasadler@deepmind.com}

\reportnumber{001} 

\renewcommand{\today}{2021-06-25}

\author[*,1]{Dan Rosenbaum}
\author[*,1]{Marta Garnelo}
\author[*,1]{Michal Zielinski}
\author[1]{Charlie Beattie}
\author[1]{Ellen Clancy}
\author[1]{Andrea Huber}
\author[1]{Pushmeet~Kohli}
\author[1]{Andrew W. Senior}
\author[1]{John Jumper}
\author[1]{Carl Doersch}
\author[*,1]{S. M. Ali Eslami}
\author[*,1]{Olaf Ronneberger}
\author[*,1]{Jonas Adler}

\affil[*]{Equal contributions}
\affil[1]{DeepMind}

\begin{abstract}
Cryo-electron microscopy (cryo-EM) has revolutionized experimental protein structure determination. Despite advances in high resolution reconstruction, a majority of cryo-EM experiments provide either a single state of the studied macromolecule, or a relatively small number of its conformations. This reduces the effectiveness of the technique for proteins with flexible regions, which are known to play a key role in protein function. Recent methods for capturing conformational heterogeneity in cryo-EM data model it in volume space, making recovery of continuous atomic structures challenging. Here we present a fully deep-learning-based approach using variational auto-encoders (VAEs) to recover a continuous distribution of atomic protein structures and poses directly from picked particle images and demonstrate its efficacy on realistic simulated data. We hope that methods built on this work will allow incorporation of stronger prior information about protein structure and enable better understanding of non-rigid protein structures.
\end{abstract}

\begin{document}
\maketitle

\section{Introduction}
Cryogenic electron microscopy (cryo-EM) using single particle analysis has become a mainstay for de novo determination of the atomic structures of proteins and other biological macromolecules, and it has allowed for the determination of the structures of many targets that were challenging or even impossible by other methods. In cryo-EM imaging,  tens of thousands to millions of highly noisy projection images of a protein are acquired using electron microscopy (the acquired images are usually called `micrographs') and a structure is determined by computationally combining information from all of them. One of cryo-EM’s advantages is simplified sample preparation, which enables structure determination of targets such as large complexes or transmembrane proteins. Another key differentiator is the ability to image flexible proteins in multiple or even a continuous distribution of configurations. Given the recent success of computational techniques in determining the static structure of proteins \citep{jumper2020}, it is likely that experimental methods which enable recovery of large spaces of conformations will be even more important moving forward.

Cryo-EM data captures such heterogeneity since each projection image shows the protein in a slightly different conformation. To make use of this inherent heterogeneity several computational methods have been proposed. The most commonly used are discrete multi-state models \citep{scheres2012relion, punjani2017cryosparc, lyumkis2013likelihood} that reconstruct a relatively small number of independent volumes (cryo-EM maps), and in a separate follow-up step fit an atomic structure of the molecule to each individual volume. Several methods have been proposed that attempt to reconstruct the cryo-EM map for the whole continuous distribution of states. These typically work by using a  continuous representation of the volume, using e.g. rotating parts of the volume \citep{nakane2018characterisation} or a low-dimensional linear subspace of volumes \citep{punjani20213d}. Each of these representations constrains the methods to recover only a relatively limited space of possible conformations.

Recently, Deep Learning (DL) techniques have shown widespread success in processing image and volumetric data and some of these are readily applied to cryo-EM, e.g. for denoising \citep{bepler2020topaz} or particle picking \citep{wagner2019sphire}. Deep learning has also been used in image reconstruction from raw data \citep{Adler2018, Arridge2019}, including as a prior for cryo-EM reconstruction \citep{kimanius2021exploiting} built into the RELION framework. Of particular interest is recent work that has demonstrated how deep neural networks can be used instead of standard voxels to represent the reconstructed volume in cryo-EM, either directly \citep{zhong2021cryodrgn} or by modelling deformations \citep{punjani20213d}.

These reconstruction algorithms have all worked in volume space, but we postulate that structural heterogeneity is better represented in atom coordinates where prior knowledge on protein structure is more readily exploited.
Movement of atoms or rigid bodies lends itself to modeling using prior knowledge about the protein physics more easily than volume displacement, especially for correlated movements. Using the atom coordinates allows us to directly recover the desired structure, without needing to go through a two step process (micrographs to volumes, then volumes to atomic models). This enables the model to share power between all states and across the entire procedure. 

\begin{figure*}[htbp]
    \centering
    \includegraphics[width=0.8\textwidth]{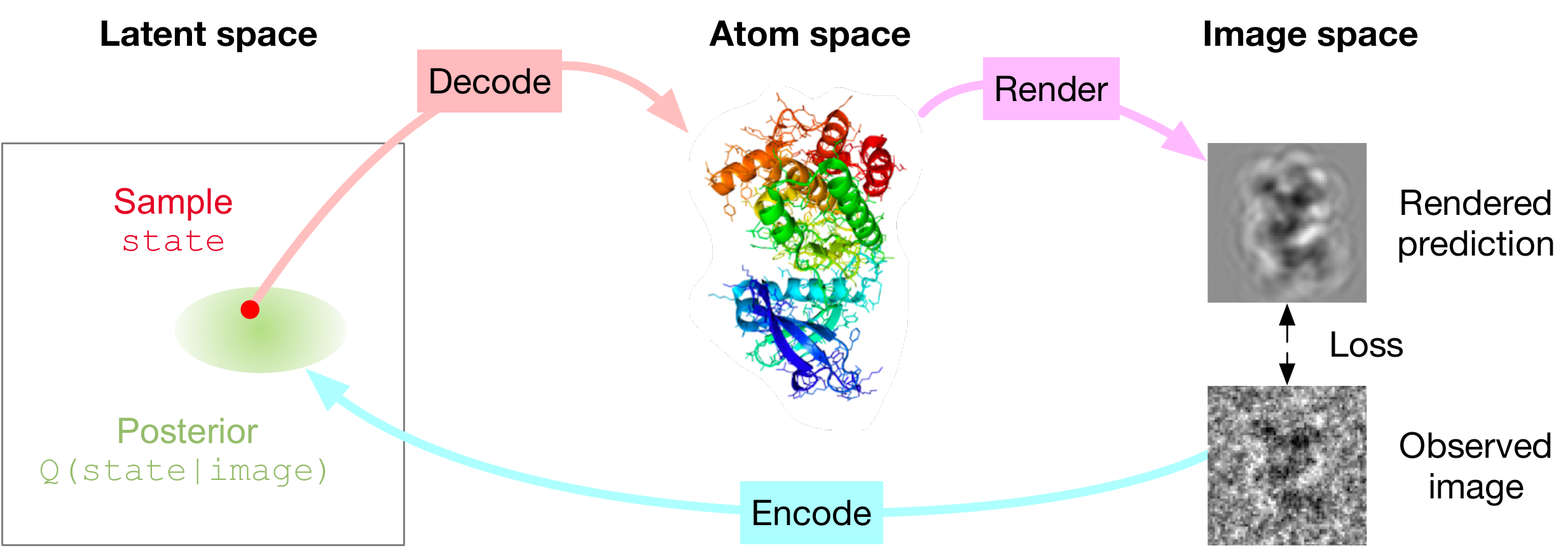}
    \caption{Overview of the proposed method. The observed image is encoded into a posterior distribution over a latent space that captures conformation and pose. Samples from this distribution are decoded into atom space and then rendered into an image that is compared with the observed image. The Render and Encode functions are only needed during training.}
    \label{fig:overview}
\end{figure*}

In this work we propose a fully end-to-end deep learning system which can reconstruct the atomic heterogeneity of proteins using deep learning, specifically Variational Auto-Encoders (VAEs) (see \autoref{fig:overview}). 
We train a model that captures the distribution of conformations, by (1) modelling the full forward process that generates an EM image: starting from latent variables that capture conformation and pose, passing through atom coordinates, and finally outputting a rendered EM image; and (2) training an inverse model (`encoder') that predicts a posterior over the latent variables given a noisy EM image.  Once the model is trained, the prior over latent space should match the accumulated posterior from all the training images, and we can use this prior to investigate the distribution of the protein conformations in atom coordinate space. We evaluate the methodology on high quality simulations and show that the method can recover multi-modality.

\section{Method}
Given a large number of EM images of a protein in different conformations and poses, our goal is to approximate the distribution of the conformations for the protein. We train a model for all cryo-EM data captured for a certain protein  and then draw samples from the trained model that correspond to different conformations. The challenge is that we have no ground truth knowledge of the conformations or poses, and the only information in the data comes from very noisy 2D projections of the protein structure that confound conformations and poses.

To tackle this, we make use of two insights: 1. We can approximate the image formation process from atom positions to images relatively well by simulating the electron microscopy process with a differentiable renderer. 2. We can formulate stereochemical constraints directly on the inferred atom positions (e.g. respecting covalent bonds).  We incorporate these insights by making some updates to the standard variational autoencoder model, which is a model that captures the distribution of images using latent random variables: We constrain the model to explicitly include the 3D position of atoms that are then rendered to the image with the differentiable renderer, and we add a regulariser directly on the atom positions.

\subsection{Background: maximum likelihood with variational autoencoders}
Variational autoencoders (VAEs) are models that capture the distribution of data, typically images, using neural networks with parameters that are trained by maximum likelihood (see \citet{doersch2021tutorial} or \citet{Kingma2019} for a detailed introduction). 

Given a dataset $Y$ of images (we use picked particles), the goal is to find the parameters of the VAE that maximise the probability of the images:
\begin{equation}
    \max_\theta \log P_\theta(Y)
\end{equation}
In order to make the distribution expressive, VAEs use a model that contains latent random variables $Z$, and separately model a prior over the latents $P(Z)$ and a conditional model $P(Y\mid Z)$ called a decoder. The decoder is usually implemented as a normal distribution with mean and variance given by $P(y\mid z)=\mathcal{N}(y ~;~ \mu_\theta(z),\sigma_\theta(z))$ where $\mu$ and $\sigma$ are deep neural networks. The downside of this is that computing the probability of a given image $y$ involves an integral over the potentially high dimensional latents:
\begin{equation}
    P_\theta(y)
    =
    \int P_\theta(y,z)dz 
    =
    \int P_\theta(y\mid z) P_\theta(z) dz
\end{equation}
While we could potentially approximate this integral by sampling $z$ according to $P(Z)$, this approach would require a very large number of samples. Standard approaches in cryo-EM \citep{scheres2012relion, punjani2017cryosparc} solve a similar problem using different techniques that work only for constrained (small) latent spaces, e.g. using exhaustive search or divide and conquer methods.

In contrast, VAEs give us a way to efficiently approximate this integral by introducing a second network with parameters $\phi$, called an encoder, that maps samples from the image space to a distribution Q in the latent space.  The core of the variational auto-encoder approach is the following equation (see \citet{Kingma2019} or \citet{doersch2021tutorial} for a derivation):
\begin{equation}
    \log P_\theta(y) - \operatorname{KL}\left[Q_\phi(z\mid y)\parallel P_\theta(z\mid y)\right]
    =
    \mathbb{E}_{z \sim Q_\phi(\cdot\mid y)} \left[\log P_\theta(y\mid z)\right] 
    - \operatorname{KL}\left[Q_\phi(z\mid y)\parallel P_\theta(z)\right]\label{eq:vae}
\end{equation}
Where KL is the Kullback-Leibler divergence.  This equation holds for any distribution $Q$ over the latent variables $z$.  Note that the KL divergence is always positive, and is zero when $Q_\phi(z\mid y)=P_\theta(z\mid y)$.  Therefore, this is a lower bound of the log-likelihood, and for sufficiently expressive $Q$ and $P$ networks, maximizing the right hand side (RHS) will result in $Q_\phi(z\mid y)$ approaching $P_\theta(z\mid y)$, and the whole expression approaching $\log P_\theta(y)$.  Therefore, optimizing the RHS will give us two things: first, we can approximately sample from $P_\theta(z\mid y)$ in an efficient way by sampling from $Q_\phi(z\mid y)$ (also called amortized inference). Second, we will also be maximizing $P_\theta(y)$ despite not having a closed-form expression for it.

In order to allow the RHS of \autoref{eq:vae} to be optimized efficiently, we implement the encoder as normal distribution parameterized by neural networks (similar to the decoder) $Q_\phi(z\mid y)=\mathcal{N}(z ~;~ \mu_\phi(y),\sigma_\phi(y))$.  To optimize, we randomly sample $y$ from the data and pass it through $Q$. $\operatorname{KL}\left[Q_\phi(z\mid y)\parallel P_\theta(z)\right]$ is a KL divergence between two normal distributions, and can be computed in closed form.  We compute $E_{z \sim Q_\phi(\cdot\mid y)} \left[\log P_\theta(y\mid z)\right]$ by sampling $z$ from $Q_\phi(z\mid y)$ and using our closed-form expression above for $P_\theta(y\mid z)$.

Note that the above procedure has the overall appearance of an autoencoder: we encode an image y via an `encoder' ($Q$), pass it through an information bottleneck ($z$), and then reconstruct $y$ as well as possible. When \autoref{eq:vae} is optimized, the prior over $Z$ matches the accumulated posterior from all images in the training set \citep{hoffman2016elbo} and the model can be used to sample new data from the same distribution by first sampling the latent $Z$ from the prior, and then using the decoder to sample $Y$ given $Z$.

\subsection{Incorporating prior knowledge in the VAE}
In our case, we are interested in a model that not only captures the distribution of images, but also captures the underlying distribution of conformations with explicit atom positions. In order to do this we update the decoder to explicitly represent the atom positions and the cryo-EM rendering process.  We do this by modelling the image distribution as a mean image plus normal distributed noise $P_\theta(y\mid z) = \mathcal{N}(y ~;~  \mu_\theta(z),\sigma_\theta(z))$,  computing the mean image as $\mu_\theta(z) =\text{render}(f_\theta(z))$, where $f_\theta (z)$ is a function containing a neural network that decodes the latents $z$ into the atom position coordinates, and `render' is the differentiable renderer that simulates the EM process and outputs an image. Our variance $\sigma_\theta (z)$ is independent of the latents $z$ and is set to correspond to the noise in the EM process. See \autoref{fig:overview} for an overview of the approach.

The benefit of constraining the VAE decoder to pass through the atom positions is twofold. First it allows us to explicitly add prior knowledge in atom space which serves as an inductive bias that constrains the space of possible image distributions. Second, it allows us to directly sample atom positions at test time, simply by running the trained decoder on samples from the prior over the latents $P(Z)$. This is in contrast to prior work that uses VAEs for cryo-EM data \citep{zhong2021cryodrgn} in image space only, without explicitly using atom positions.

\subsection{Model components}
\begin{figure}[htbp]
    \centering
    \includegraphics[width=\textwidth]{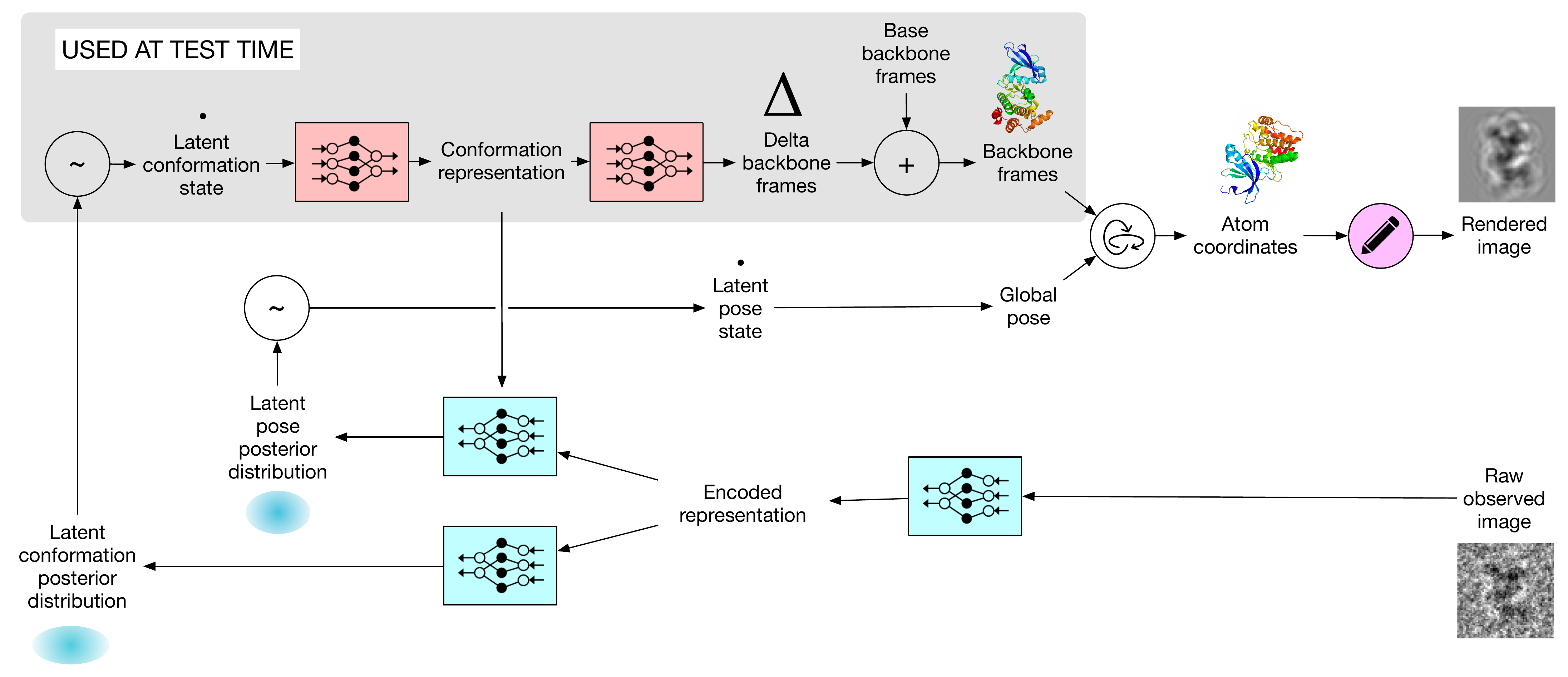}
    \caption{Details of the proposed architecture. Arrows indicate the information flow. Trainable network blocks are represented as colored boxes, static operations are represented as circles. }
    \label{fig:architecture}
\end{figure}

We now describe the different model components as depicted in \autoref{fig:architecture}. 
The two main components with trainable parameters are the encoder (colour coded in blue) and the decoder (colour coded in red). See \autoref{sec:appendix} for more exact details.

\paragraph{The encoder} is split into three parts, 1. A neural network that takes the observed image and encodes it into a general representation vector, 2. A network that encodes the representation vector into a posterior distribution over the conformation latent variable and 3. A network that encodes the representation vector into a posterior distribution over the pose latent variable. Splitting the latent variables into two parts that correspond to the conformation and pose, gives us more flexibility in controlling these two confounding factors.  One way we do this is by conditioning the pose encoder on samples of the conformation, allowing the model to capture a multi-modal distribution more easily (essentially this is an auto-regressive posterior). 

\paragraph{The decoder} consists of a neural network that takes the sampled latents and outputs a delta of the conformation relative to some base conformation. The pose latents are directly used as a global 2D translation and 3D rotation, where the latter is transformed into a rotation matrix using Gram-Schmidt orthogonalization (see \autoref{sec:appendix} for details).

\noindent The rest of the model components are fixed and do not contain trainable parameters. 

\paragraph{Protein structure.} We define all conformations relative to a base conformation which can be obtained by a conventional single state reconstruction, or, e.g., by computational modelling (foregoing the need for obtaining an experimental structure or template). In the presented experiments we simply use a single state conformation from the Protein Data Bank. We found that using some base conformation was critical -- randomly initialising the residues did not give useful structures -- but the exact choice did not seem to matter. The final conformation is obtained by predicting a relative translation and rotation for each residue (termed ‘delta backbone frames’ in \autoref{fig:architecture}). Each residue (including its sidechain) is represented as an individual rigid body \citep{jumper2020}, the inter-residue bonds along the backbone are only indirectly constrained by the backbone continuity loss (see below).

\paragraph{Renderer.} In order to compute the data log likelihood of a predicted atom configuration we pass it to a renderer that predicts the mean image, on top of which we model the noise as a normal distribution.  The exact forward operator in cryo-EM is relatively intricate (see \citet{vulovic2013image} for an in depth description). However it is usually assumed that it can be modelled as a projection of the electron density followed by a convolution with the Contrast Transfer Function (CTF).

We model the electron density for each heavy atom by a single Gaussian blob with standard deviation 1Å and unit mass. This can be easily extended to more accurate formulations as needed. Further, note that our input data was generated using a much more complex model that we describe later.

By using Gaussians we can compute the rendering function analytically by projecting all Gaussians to each of the output image pixels directly, without passing through a voxel representation that would incur a loss of precision. This can be done efficiently using the separability of the Gaussian kernels, and by computing the corresponding values for the x and y axes separately. This projection operator is also end-to-end differentiable, which we need for training. We represent the CTF directly in real space as a convolution and apply it to the projected image using the fast Fourier transform with padding. This too is fast and end-to-end differentiable.

\subsection{Training}
\begin{figure}[htbp]
    \centering
    \includegraphics[width=\textwidth]{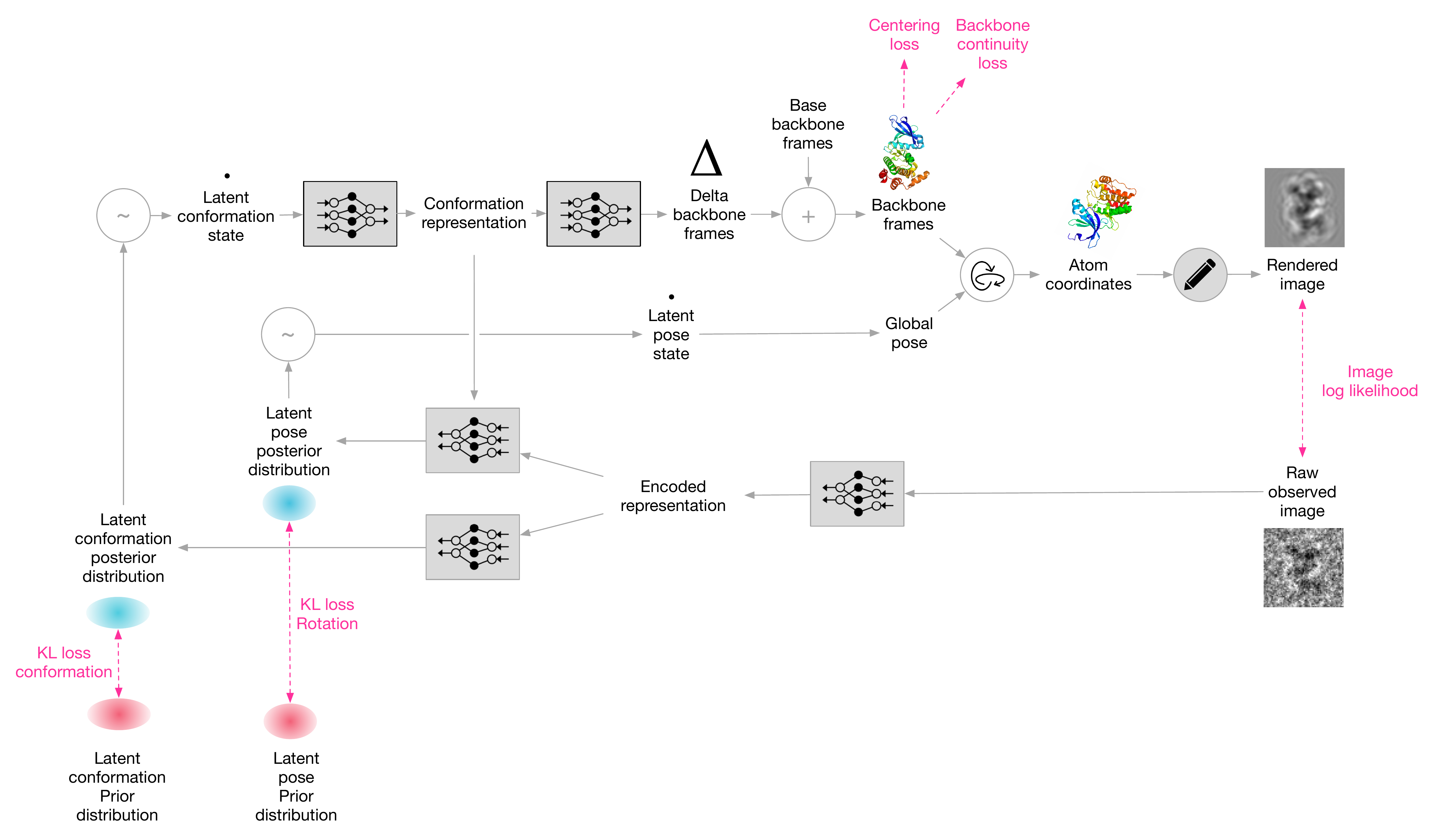}
    \caption{Losses used to train the model.}
    \label{fig:losses}
\end{figure}
To train the VAE, the image log likelihood and the KL terms need to be optimized (\autoref{fig:losses}).
In addition to the loss computed from reconstructing images, we add an auxiliary loss term in atom space that consists of the mean squared deviation from the expected bond lengths along the backbone (termed `Backbone continuity loss' in \autoref{fig:losses}). This loss helps significantly in keeping the predictions physically plausible, but it corresponds to a simple prior and our model could also incorporate much stronger priors on protein structure as needed. We also include a small squared error loss (\autoref{fig:losses} `Centering loss') on the center of mass to keep the structure centered around zero, forcing the network to represent translations using the global pose rather than moving all residues independently.

Since the encoded posterior distribution can be too simple to capture the complex multimodal distribution of conformation and pose given a single noisy image, we make a modification to standard VAE training, by drawing multiple samples from the posterior and \emph{summing} their corresponding likelihoods (the standard VAE training objective \emph{multiplies} the likelihoods of the samples by adding their log-likelihoods). This is a similar objective to IWAE \citep{burda2015IWAE} except that we do this only for the reconstruction term without the KL term, therefore our objective is no longer a proper bound on the log-likelihood. We find that this method encourages the model to explore more modes while training, most likely because the model does not get penalized that strongly for producing some low-probability samples. Without this IWAE-like objective our models did not train successfully. We also weight the contribution of the KL with a hyper parameter $\beta$,  similar to beta-VAEs \citep{higgins2016beta},  in order to better control the tradeoffs between reconstruction error and prior/posterior matching. Including the structure loss $\mathcal{L}_\text{structure}$, the objective per image $y_i$ is:
\begin{equation}
    -\log\left[ \frac{1}{N}\sum_{n=1}^N P_\theta(y_i\mid z_n)\right]
    + \beta\operatorname{KL}\left[Q_\phi(z\mid y_i)\parallel P_\theta (z)\right]
    + \frac{1}{N}\sum_{n=1}^N \mathcal{L}_\text{structure}(f_\theta(z_n))
\end{equation}
Where $z_1 \dots z_N$ are sampled from the posterior $P_\theta(z\mid y_i)$ and $f(z)$ is the atom structure decoded from the latent $z$. The prior over the latents $P_\theta(z)$ is set to be a standard normal distribution, which for the global pose corresponds to a normal distribution of translation around the center, and a uniform distribution of rotation.  We optimize the average of this loss over all images using stochastic gradient descent with mini-batches (Adam~\citep{Kingma2015}). 

We find that in order to successfully train the model, an initial phase of pose-only training is needed. Therefore we first train the model by predicting only poses of the base conformation using 128 samples of the pose latents. After convergence we also start to predict the conformation deltas using 64 samples of the conformation latents, where each sample is used with 4 pose latent samples (total of 256 samples from each posterior). We use mini-batches of size 256.

For evaluation we draw unconditional samples from the prior on conformation latents, and we do not need to deal with global poses at all.  The process is therefore to draw a number of samples from the prior $P(Z)$, then decode them into delta backbone frames, and apply them to the base backbone frames, resulting in different conformations of the protein.

\section{Simulated dataset}
In order to accurately measure progress and determine if we successfully recovered the right distribution, we decided to apply our method in a simulation study. However, we did try to make our simulations as realistic as possible, e.g. by taking the conformations from an externally validated Molecular Dynamics (MD) simulation and simulating the image formation process using a high quality simulator including e.g. contrast transfer function, realistic noise models and solvation.

Externally validated MD simulations are not abundant, and we selected conformations from a  trajectory of Aurora A Kinase (AurA) in an apo (unbound) state, generated with MD simulations by Folding@Home \citep{cyphers2017water} as an interesting model system. AurA is a non-membrane monomeric enzyme with flexible catalytic and activation loops.

\begin{figure}[htbp]
    \centering
    \sf\footnotesize
    \raisebox{2.5cm}{\large \textbf{a}}~
    \begin{tabular}{lc}
        \toprule
        \textbf{Parameter} & \textbf{Value}\\
        \midrule
        Number of micrographs & 3000 \\
        Number of picked particles & 63,000\\
        Dose & 1000 e/Å$^2$ \\
        Image size & 64 px $\times$ 64 px\\
        Image pixel size & 1.2 Å\\
        Magnification & 25,000\\
        Defocus & 5,000 Å\\
        Accelerating voltage & 300 keV\\
        Spherical aberration (CS) & 2.1nm\\
        Amplitude contrast ratio & 0.06\\
        \bottomrule
    \end{tabular}
    ~~~~\raisebox{2.5cm}{\large \textbf{b}}
    \begin{tabular}{l}
        \includegraphics[width=0.35\textwidth]{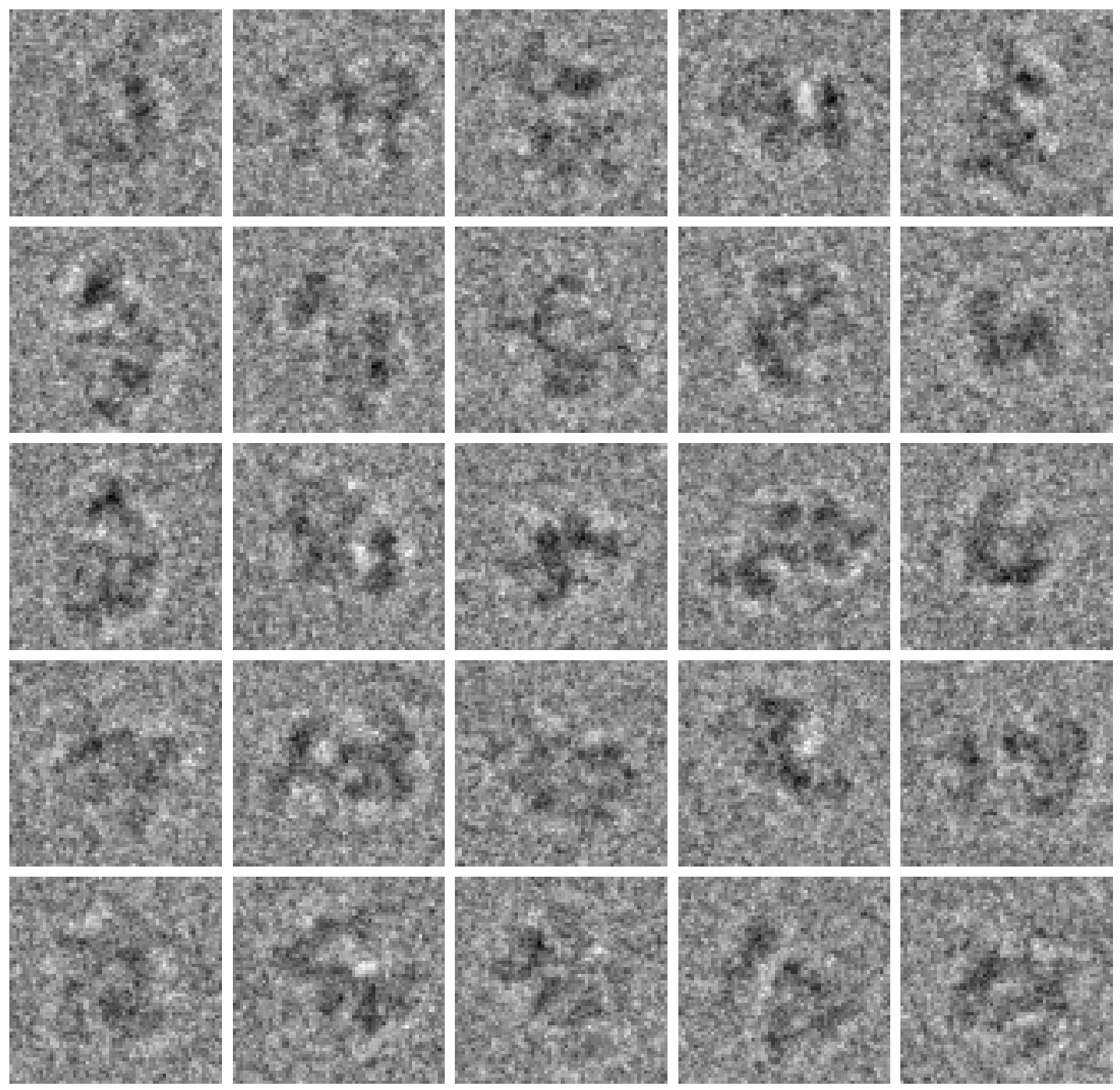}
    \end{tabular}
    \caption{Cryo-EM simulation. (\textbf{a}) Parameters for cryo-EM simulator. (\textbf{b}) Sample of picked particles images from our simulated dataset.}
    \label{fig:simulation} 
\end{figure}

The 3000 sampled states served as the input to simulate the cryo-EM imaging process, using TEM-simulator \citep{rullgaard2011simulation} to realistically model the image formation and noise of a real cryo-electron microscope. Each sample is placed randomly (rotation and translation) on the resulting micrograph. Given limited options of highly-dynamic MD on large proteins, we selected a protein whose size (33 kDa, 282 residues) is below current practical experimental limits of cryo-EM. Therefore we applied a $\sim$10 times higher-than-usual electron dose ($\sqrt{10}\sim 3$ times higher SNR) to make the difficulty of determining the rotations roughly comparable to a protein of about 1000 residues. Otherwise, our simulation setup was chosen to broadly follow standard practice. Details of parameters used for the simulator can be found in \autoref{fig:simulation}a. Using known positions of particle placements we extracted and shuffled particles from the resulting micrograph. A random sample of images used as the input to the model can be seen in \autoref{fig:simulation}b.

\section{Evaluation}
\begin{figure}[htbp]
    \centering
    \includegraphics[width=\textwidth]{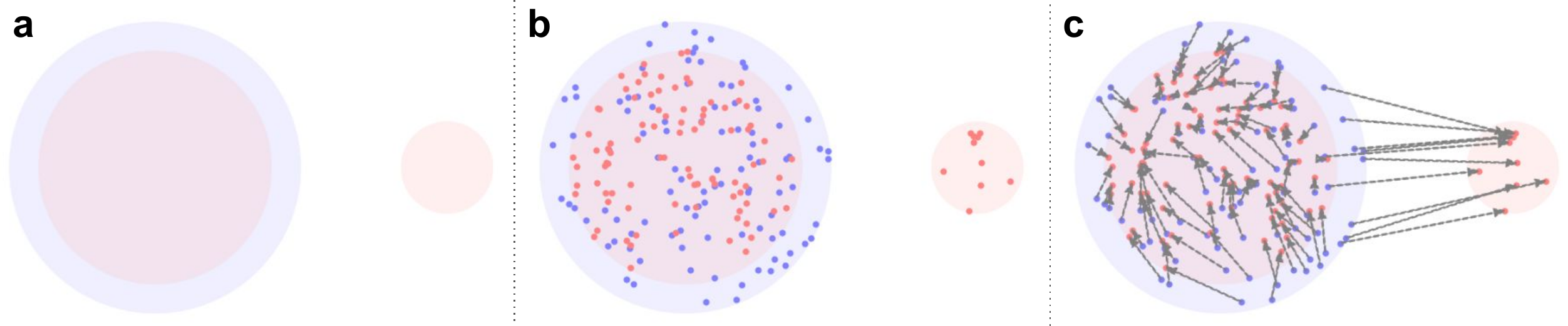}
    \caption{Approximation of the earth mover’s distance between two distributions. (\textbf{a}) Predicted (bi-modal) distribution (red) and ground truth distribution (blue) represented by colored regions in a 2D space. (\textbf{b}) Dots represent random samples drawn from both distributions. (\textbf{c}) Arrows indicate the optimal one-to-one assignments of the samples found by the Hungarian algorithm. }
    \label{fig:EM-RMSD} 
\end{figure}

In classical structure determination the result is a single structure, and the quality of a prediction or reconstruction can be measured by comparing it to a ground truth structure using e.g. the root mean squared deviation (RMSD). In our setting both the result and ground truth are given by distributions of structures and to accurately assess our model we need an evaluation metric that can compare two arbitrary distributions based on samples from them. To do this, we approximate the earth mover’s distance (EMD), also known as Wasserstein-1 metric \citep{villani2009optimal},  between these two distributions (\autoref{fig:EM-RMSD}a).

We denote our specific metric of choice as Earth Mover’s Distance based on Root Mean Squared Deviation (EMD-RMSD).  We start by drawing $N$ samples from ground truth distributions and N samples from the model (\autoref{fig:EM-RMSD}b) (in particular we choose $N=2048$). We then compute the RMSD between all prediction-ground truth pairs after rigidly aligning the two protein structures together. Given an $N\times N$ matrix of RMSD distances we compute optimal linear one-to-one assignment between the ground truth samples and the model samples using the Hungarian algorithm (\autoref{fig:EM-RMSD}c) (i.e. each sample prediction gets assigned to a close ground truth state, to minimize the global assignment cost). Computed RMSDs for the assigned model-ground truth pairs are averaged across pairs and finally reported as the evaluation metric.

\section{Results}
\begin{figure}[htbp]
    \centering
    \includegraphics[width=\textwidth]{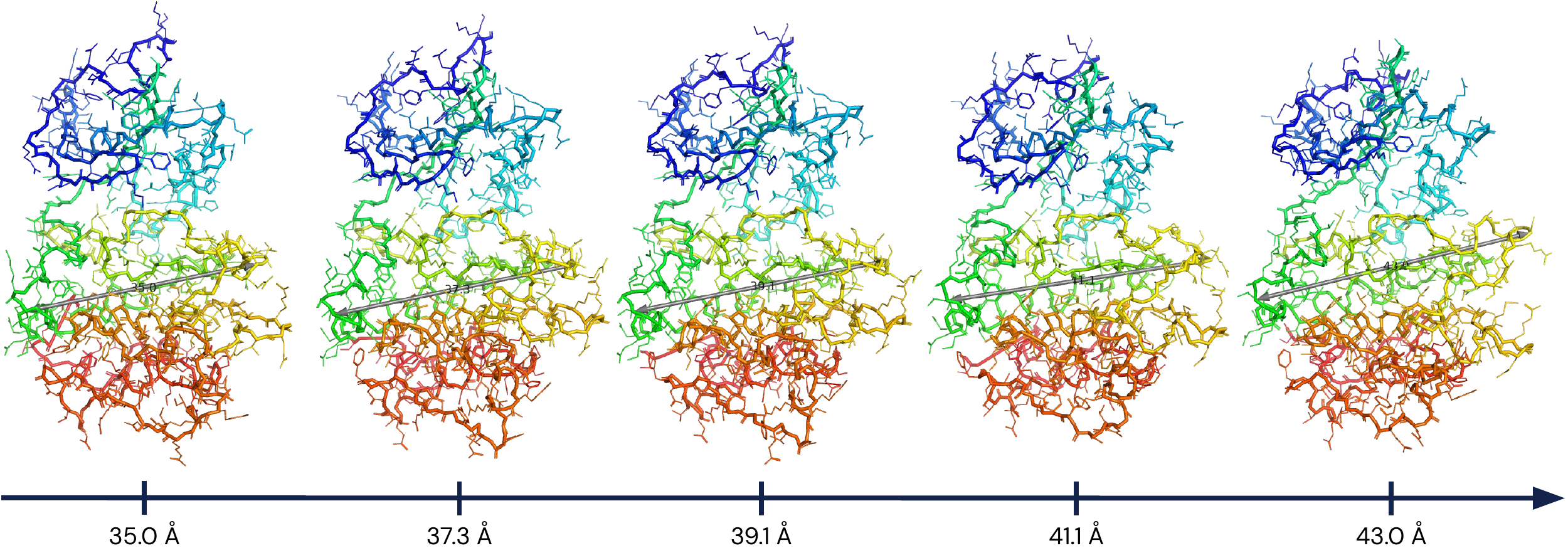}
    \caption{Latent space interpolation. Distances plotted correspond to T288C$\alpha$--R255C$\zeta$, also indicated by the grey double-arrow superimposed on the protein structure.}
    \label{fig:latent space interpolation} 
\end{figure}

We were able to successfully train a neural network to represent diverse states of the AurA protein from 2D particle images. According to our EMD-RMSD metric we fit the true distribution to within 3.35 Angstrom. \autoref{fig:latent space interpolation} 
shows five samples in a latent space interpolation demonstrating the continuous change between two states obtained by sampling $z$ along a line in latent space. We note that the reconstructed proteins look qualitatively correct with no major steric clashes or bond breaking. However, upon closer inspection there seems to be some issues such as broken H-bonds in the beta-sheets (see supplementary \href{https://youtu.be/5Kpf6gaFhVQ}{video 1}).

\begin{figure}[htbp]
    \centering
    \includegraphics[width=0.6\textwidth]{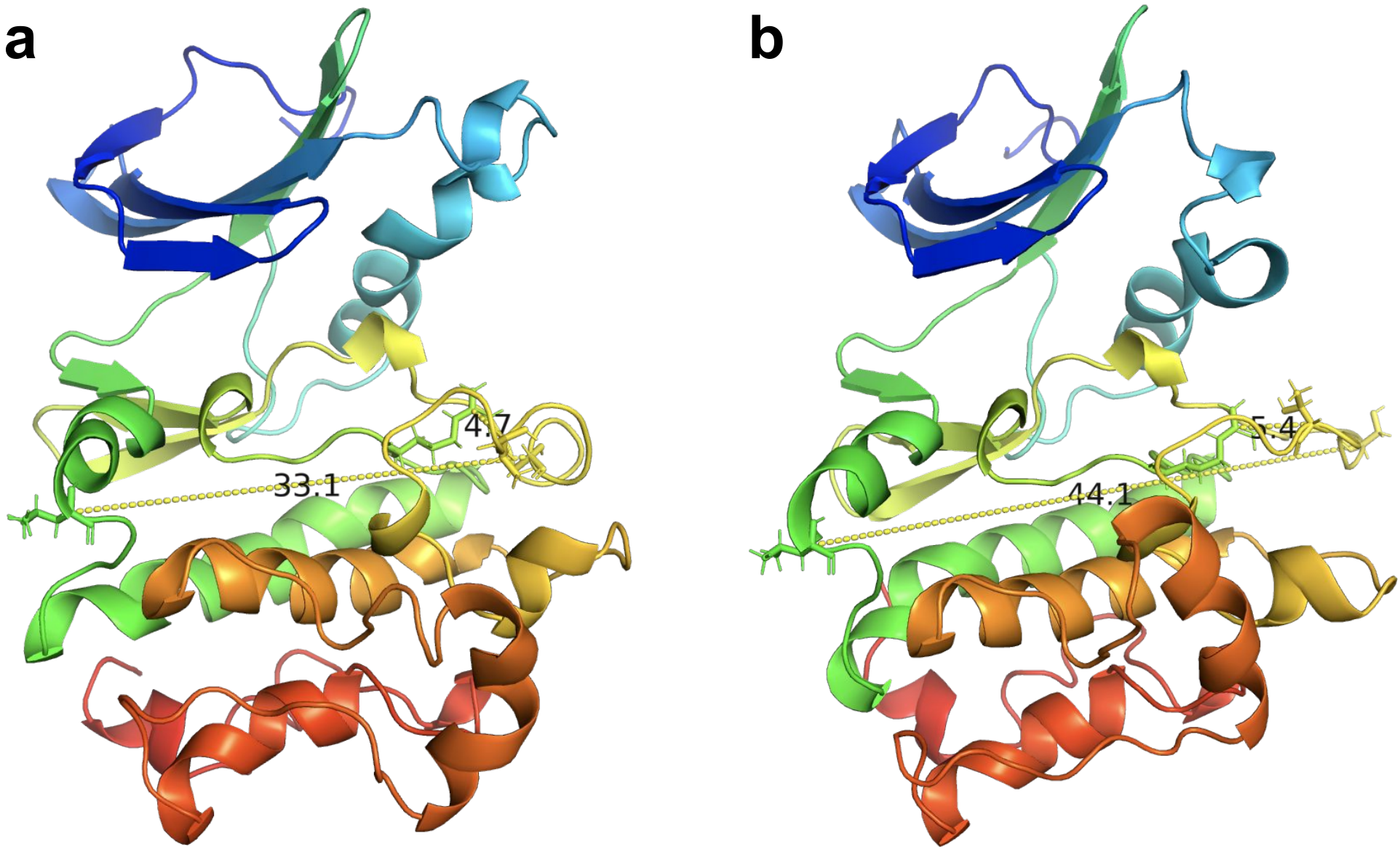}
    \caption{Visualisation of ground truth conformations in the AurA protein from MD simulations. Yellow dashed lines show the distances S284C$\alpha$--L225C$\alpha$ and T288C$\alpha$--R255C$\zeta$. (\textbf{a}) Short state. (\textbf{b}) Long state.}
    \label{fig:md conformations} 
\end{figure}

To investigate how well we have recovered the set of all states present in the data we investigated marginal distributions, specifically pairwise distances analysed previously in \citet{ruff2018dynamic}. The two distance pairs we studied are S284C$\alpha$--L225C$\alpha$ (tip of the activation loop and the distal surface of the kinase domain) and  T288C$\alpha$--R255C$\zeta$  (phosphorylated site in the activation loop and HRD motif in the catalytic loop). These distances are visualised in \autoref{fig:md conformations} 
for long and short states in the AurA protein. Such distances between pairs of atoms/residues are pose invariant and are also routinely used to capture the conformations of the protein using experimental techniques such as FRET \citep{algar2019fret}.

\begin{figure}[htbp]
    \centering
    \includegraphics[width=0.8\textwidth]{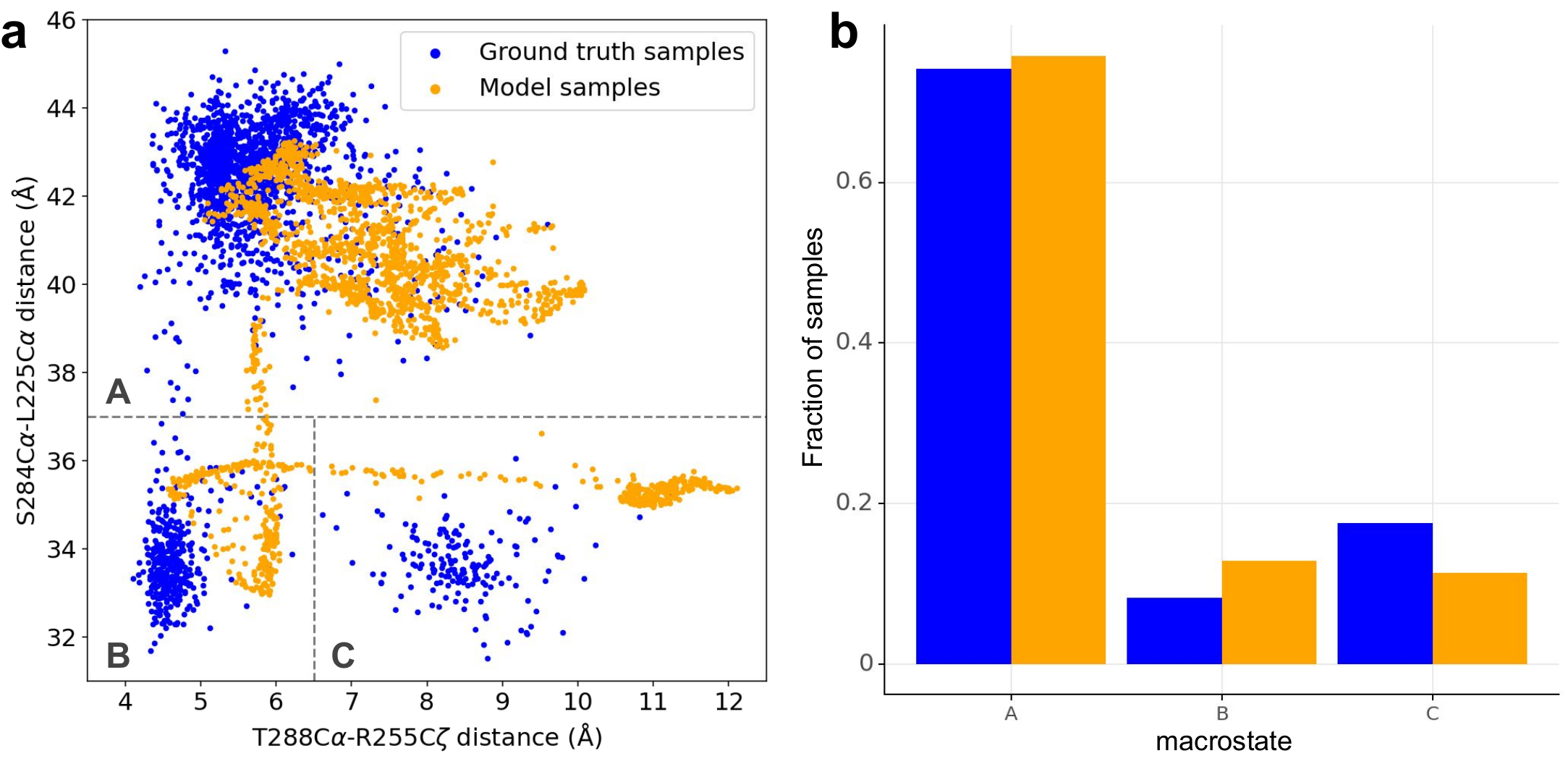}
    \caption{Predicted distribution vs. ground truth distribution. (\textbf{a}) Illustration of sample quality with each 3D protein structure represented as a point defined by two pairwise distances. Dashed lines stratify samples into groups using 6.5 Å as a cutoff on the X-axis and 37 Å as a cutoff on the Y-axis. (\textbf{b}) Proportion of samples in each of the groups.}
    \label{fig:distance distributions} 
\end{figure}
In \autoref{fig:distance distributions} we show the distribution of the ground truth MD samples compared to distributions reconstructed with the model over two such distances. 
We note that we have successfully recovered all three major modes seen in this projection, including correctly determining their approximate relative distributions. However there are some noticeable failures. The first is that our method seems to find biased distances. Another is that the latent space is not as smooth as could be hoped, rather showing a certain degenerate sheet-like structure.

\begin{figure}[htbp]
    \centering
    \includegraphics[width=0.8\textwidth]{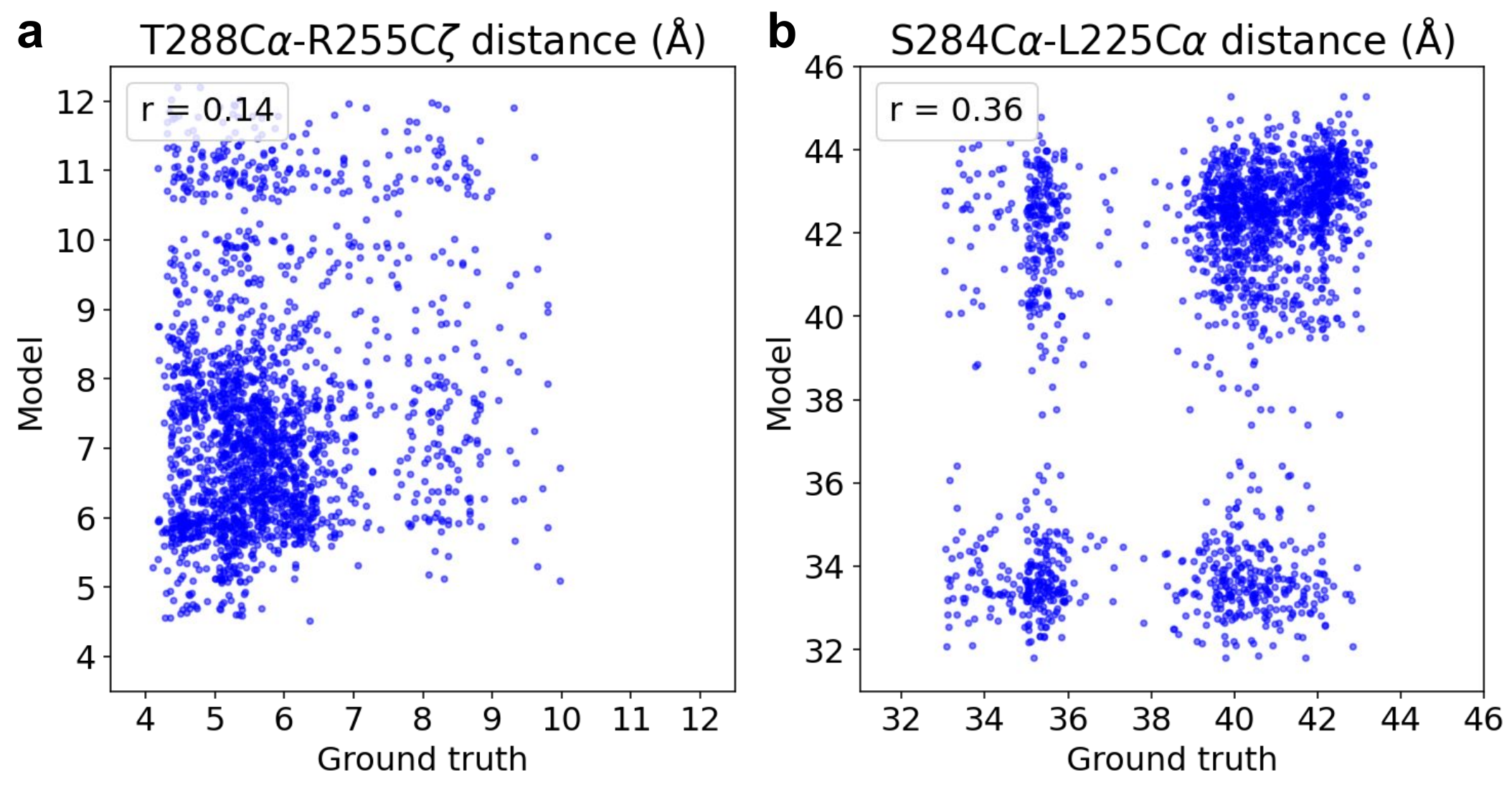}
    \caption{Conditional model prediction visualised by pairwise distances. A perfect reconstruction would give diagonal samples, but due to high noise we observe only a minor correlation (shown as Pearson correlation coefficient). (\textbf{a})~Distance T288C$\alpha$--R255C$\zeta$, correlation r=0.14 (\textbf{b})~Distance S284C$\alpha$--L225C$\alpha$, correlation  r=0.36}
    \label{fig:conditional prediction} 
\end{figure}

While the unconditional samples roughly recover the desired distribution, when sampling conditionally (i.e., sampling from the posterior) we observe that the conditional samples seem relatively uncorrelated to the true distributions for the sample that we condition on (\autoref{fig:conditional prediction}).
This indicates that while the noise level in the input 2D picked particles is so high that no particle contains sufficient data for a useful reconstruction on its own, our model is able to share capacity across the tens of thousands of data points and still recover reasonable unconditional distributions.

\section{Discussion}
One important finding of our experiments was that a standard VAE combined with a renderer is not able to learn the distribution of conformations and poses from cryo-EM images. The key contributions to make this end-to-end approach work were the disentangling of the distributions over poses and conformations, the prediction of the structures relative to a base structure, training with multiple samples for each observed image (IWAE style) and relatively large batch sizes. 

The disentanglement of the poses and conformations seems to be especially crucial. Architectures with a single latent space for encoding both conformation and pose tended to converge to a degenerate solution in our experiments. The model explained all variations in the observed images by creating a large number of different degenerate 3D structures that produce a good image in a single pose only, and never reused these structures in a different pose to explain other observed images. The disentanglement into two latent spaces and the pre-training of the pose prediction using the fixed base structure help significantly to produce non-degenerated structures. We assume that the major challenge in this setup with a renderer in the decoder is the very fragmented loss space. The gradients in the image space only map (through the differentiable renderer) into meaningful gradients in the atom space if the predicted structure and pose is sufficiently close to the true structure and pose. In a randomly initialized network the large majority of the gradients will point into completely wrong directions and result in many degenerate structures. A network that is pre-trained to predict the pose and therefore creates predictions that are already sufficiently close to the correct answer results in a large fraction of gradients that point in the correct direction, and leads to a successful training via gradient descent.

The other major challenge is the extremely low signal-to-noise ratio in the Cryo-EM images, which means that it will never be possible to predict the exact pose or conformation of the real-world protein that is depicted in a single image – we can only narrow it down to a distribution of likely conformations and a distribution of likely poses. Our training procedure addresses this inherent uncertainty by predicting an ensemble of possible images for each observed image, and by computing a weighted loss over them. We believe that further experimentation with generative modelling objectives and techniques (e.g. wake-sleep style training \citep{hinton1995wake}) can lead to improvements, and that performance is far from saturated.

One of the main limitations of our study is the evaluation on simulated data only. We have taken significant steps to set up a highly accurate simulation, both with respect to structural diversity of the samples but also with respect to the physics of image formation. However, we fully acknowledge that simulations do not account for all aspects of the cryo-EM experiments (from contaminants to particle degradation) and further translational work is required to prove the viability of the presented methods for experimental data. A further limitation is that the current method assumes the atomic composition to be known and fixed. Finally, more work is needed in order to exactly model the side chains.

While we used relatively little prior knowledge of protein structure outside the covalent bonds in this study, one of the most exciting prospects is that we now have a setting in which adding further prior knowledge about the atomic structure of proteins is straightforward. Given the recent success of using machine learning for protein structure prediction \citep{senior2019protein,jumper2020} we expect significant further improvement from incorporating such prior information into this framework.

\section{Conclusion}
In this article we proposed a fully end-to-end deep-learning-based scheme for the reconstruction of distributions of atomic protein structures from cryo-EM experimental data working directly from picked particles. The scheme uses variational autoencoders, not only to represent the diversity of states but also to handle the unknown rotation and translation that are inherent to data obtained with cryo-EM.

While the presented results are constrained to simulated datasets, we think the method’s success indicates that deep learning techniques could allow the end-to-end reconstruction of atomic diversity directly from measured data and gives a way to directly combine experimental data with very strong priors from machine learning. We hope that both of these directions will be explored in future studies.

\section{Acknowledgements}
The authors would like to thank Andrew Zisserman, Clemens Meyer, Ellen Zhong, Oriol Vinyals, Rob Fergus, and Simon Kohl for their support and insightful discussions. 

\section{Appendix}\label{sec:appendix}
\begin{figure}[htbp]
    \centering
    \includegraphics[width=\textwidth]{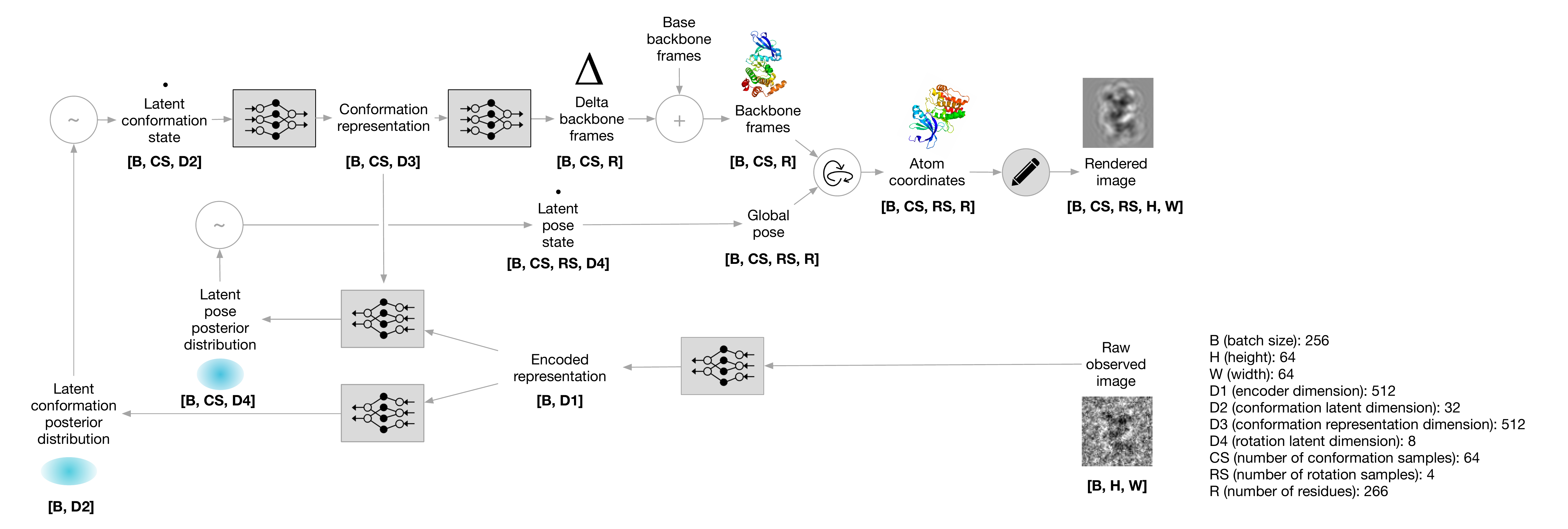}
    \caption{Shapes of the arrays of the model.}
    \label{fig:shapes} 
\end{figure}

In the following section we give all the details of our model and training setup that we believe are needed for implementation. See \autoref{fig:shapes} for a diagram that contains all the shapes of the model’s representations. 

The first step of our encoding is given by a relatively small neural network. We flatten the input image $y$ ($64\times 64$ pixels) to a 4096-dimensional vector, apply layer normalization \citep{ba2016layer} to get unit scaling, and then apply a four layer multilayer perceptron (MLP) with widths 2048, 1024, 512, 512 and ReLU nonlinearities. MLP's worked in our experiments much better than convolutional networks. We assume that the ability of MLP's to ``see'' all pixels at once helps to deal with the extreme noise in the images.
We use the resulting  representation for defining the conformation latent and the pose latent.

For the conformation latent we apply layer norm and a linear layer to get a 32-dimensional vector $\mu_\text{conf}(y)$ and another linear layer with a softplus activation to get a 32-dimensional vector $\sigma_\text{conf}(y)$. We use these vectors to define our distribution $Q(z_\text{conf}\mid y)=\mathcal{N}(z_\text{conf} ~;~ \mu_\text{conf}(y), \sigma_\text{conf}(y)).$

Next we draw 64 samples from this distribution and decode them independently. For each sample $z$, we apply a small residual network (ResNet) of width 256 with 5 blocks consisting of layer norm, linear layer, leaky ReLU, linear layer. Finally, we apply layer norm to the final output and a linear layer with width 512 which gives the decoded representation. This decoded representation is then fed to a final linear layer with width $9 \cdot N_\text{residues}$ which we use to compute the poses of all atoms in the protein. Each 9D vector is used to compute the translation and rotation of one residue (‘frame’), by interpreting it as a 3D translation vector $\vec{t} \in \R^3$, and two 3D rotation vectors $\vec{v}_1,\vec{v}_2 \in \R^3$ that are orthogonalized by Gram-Schmidt into a full 3x3 rotation matrix $R \in \R^{3\times3}$:
\begin{align*}
    \vec{e}_1 &= \vec{v}_1 / \norm{\vec{v}_1}\\
    \vec{u}_2 &= \vec{v}_2 - \vec{e}_1 \left(\vec{e}_1\vec{v}_2\right)\\
    \vec{e}_2 &= \vec{u}_2 / \norm{\vec{u}_2}\\
    \vec{e}_3 &= \vec{e}_1 \times \vec{e}_2\\
    R &= \left(\vec{e}_1,\vec{e}_2,\vec{e}_3\right)
\end{align*}

The rotations and translations are then applied to the base backbone frames.  

Using the encoded representation of the observed image and the decoded representation of the sampled 64 conformations, we next compute a posterior distribution of possible global poses for each conformation. To do so, we start by applying a layer norm to the encoded representation and then a linear layer. We add the state decoder representations to this and run a small 3 hidden layer, width 32, MLP on top of it. Finally we compute 8-dimensional vectors $\mu_\text{pose}(y,z_\text{conf})$, and $\sigma_\text{pose}(y,z_\text{conf})$ (like for the conformations) which define our distribution $Q(z_\text{pose} \mid y, z_\text{conf}) = \mathcal{N}(z_\text{pose} ~;~ \mu_\text{pose}(y,z_\text{conf}), \sigma_\text{pose}(y,z_\text{conf}))$. We sample from this distribution to get the pose latents, which convert to a global pose (translation and rotation) by interpreting the first two coordinates as a 2D translation $\vec{t} \in \R^2$ in the detector plane (ignoring translations along the beam direction) and the final 6 coordinates as two 3D rotation vectors $\vec{v}_1,\vec{v}_2 \in \R^3$ that are then orthonormalized using Gram-Schmidt to obtain the global rotation matrix. Finally, we rotate and translate each structure in the ensemble according to the corresponding global pose.

We trained our model in two stages. In the initial stage we only train the pose encoder/decoder using the fixed base structure only. This training was done for about 2000 epochs. In the second stage we train the entire network for about 500 epochs. For both stages we use mini-batches containing 256 observed images. In total, training took a few days to converge. Due to run-to-run variance, we trained three models independently, using different random seeds, and selected the one with highest log likelihood.

We implemented all components of the network in JAX \citep{jax2018github} using Haiku \citep{haiku2020github}, jaxline \citep{deepmind2020jax} and Optax \citep{optax2020github} and trained the networks using 8 TPUv3 devices. We used the Adam optimizer with a base learning rate of $3\cdot 10^{-4}$. In the first stage a learning rate schedule reduced the learning rate by a factor of 0.5 after 1000 epochs. The weights for the individual losses were $\beta = 1$ for the KL loss in the conformation latent space, $\beta = 0.01$ for the KL loss in the pose latent space, 0.1 for the centering loss, and 0.01 for the backbone continuity loss.

We use truncated normal initialisation (scaled by $1/\sqrt{N_\text{fan in}}$, following \citep{ioffe2015batch}), and zero initialise all residual updates.

\section{Supplementary Material}

Supplementary video 1 (available as ancillary file in arXiv, and in \url{https://youtu.be/5Kpf6gaFhVQ}) shows the conformations from latent space interpolation (similar to \autoref{fig:latent space interpolation}).

\bibliography{main}

\begin{thebibliography}{32}
\providecommand{\natexlab}[1]{#1}
\providecommand{\url}[1]{\texttt{#1}}
\expandafter\ifx\csname urlstyle\endcsname\relax
  \providecommand{\doi}[1]{doi: #1}\else
  \providecommand{\doi}{doi: \begingroup \urlstyle{rm}\Url}\fi

\bibitem[Adler and \"Oktem(2018)]{Adler2018}
J.~Adler and O.~\"Oktem.
\newblock Learned primal-dual reconstruction.
\newblock \emph{IEEE Transactions on Medical Imaging}, 37\penalty0
  (6):\penalty0 1322--1332, 2018.
\newblock \doi{10.1109/TMI.2018.2799231}.

\bibitem[Algar et~al.(2019)Algar, Hildebrandt, Vogel, and
  Medintz]{algar2019fret}
W.~R. Algar, N.~Hildebrandt, S.~S. Vogel, and I.~L. Medintz.
\newblock Fret as a biomolecular research tool—understanding its potential
  while avoiding pitfalls.
\newblock \emph{Nature methods}, 16\penalty0 (9):\penalty0 815--829, 2019.

\bibitem[Arridge et~al.(2019)Arridge, Maass, \"Oktem, and
  Sch\"onlieb]{Arridge2019}
S.~Arridge, P.~Maass, O.~\"Oktem, and C.-B. Sch\"onlieb.
\newblock Solving inverse problems using data-driven models.
\newblock \emph{Acta Numerica}, 28:\penalty0 1–174, 2019.
\newblock \doi{10.1017/S0962492919000059}.

\bibitem[Ba et~al.(2016)Ba, Kiros, and Hinton]{ba2016layer}
J.~L. Ba, J.~R. Kiros, and G.~E. Hinton.
\newblock Layer normalization.
\newblock In \emph{NeurIPS 2016 Deep Learning Symposium}, 2016.
\newblock URL \url{https://arxiv.org/abs/1607.06450}.

\bibitem[Babuschkin et~al.(2020)Babuschkin, Baumli, Bell, Bhupatiraju, Bruce,
  Buchlovsky, Budden, Cai, Clark, Danihelka, Fantacci, Godwin, Jones, Hennigan,
  Hessel, Kapturowski, Keck, Kemaev, King, Martens, Mikulik, Norman, Quan,
  Papamakarios, Ring, Ruiz, Sanchez, Schneider, Sezener, Spencer, Srinivasan,
  Stokowiec, and Viola]{deepmind2020jax}
I.~Babuschkin, K.~Baumli, A.~Bell, S.~Bhupatiraju, J.~Bruce, P.~Buchlovsky,
  D.~Budden, T.~Cai, A.~Clark, I.~Danihelka, C.~Fantacci, J.~Godwin, C.~Jones,
  T.~Hennigan, M.~Hessel, S.~Kapturowski, T.~Keck, I.~Kemaev, M.~King,
  L.~Martens, V.~Mikulik, T.~Norman, J.~Quan, G.~Papamakarios, R.~Ring,
  F.~Ruiz, A.~Sanchez, R.~Schneider, E.~Sezener, S.~Spencer, S.~Srinivasan,
  W.~Stokowiec, and F.~Viola.
\newblock The {D}eep{M}ind {JAX} {E}cosystem, 2020.
\newblock URL \url{http://github.com/deepmind}.

\bibitem[Bepler et~al.(2020)Bepler, Kelley, Noble, and Berger]{bepler2020topaz}
T.~Bepler, K.~Kelley, A.~J. Noble, and B.~Berger.
\newblock Topaz-denoise: general deep denoising models for cryoem and cryoet.
\newblock \emph{Nature communications}, 11\penalty0 (1):\penalty0 1--12, 2020.

\bibitem[Bradbury et~al.(2018)Bradbury, Frostig, Hawkins, Johnson, Leary,
  Maclaurin, Necula, Paszke, Vander{P}las, Wanderman-{M}ilne, and
  Zhang]{jax2018github}
J.~Bradbury, R.~Frostig, P.~Hawkins, M.~J. Johnson, C.~Leary, D.~Maclaurin,
  G.~Necula, A.~Paszke, J.~Vander{P}las, S.~Wanderman-{M}ilne, and Q.~Zhang.
\newblock {JAX}: composable transformations of {P}ython+{N}um{P}y programs,
  2018.
\newblock URL \url{http://github.com/google/jax}.

\bibitem[Burda et~al.(2016)Burda, Grosse, and Salakhutdinov]{burda2015IWAE}
Y.~Burda, R.~Grosse, and R.~Salakhutdinov.
\newblock Importance weighted autoencoders.
\newblock In \emph{Proceedings of the International Conference on Learning
  Representations}, 2016.

\bibitem[Cyphers et~al.(2017)Cyphers, Ruff, Behr, Chodera, and
  Levinson]{cyphers2017water}
S.~Cyphers, E.~F. Ruff, J.~M. Behr, J.~D. Chodera, and N.~M. Levinson.
\newblock A water-mediated allosteric network governs activation of aurora
  kinase a.
\newblock \emph{Nature chemical biology}, 13\penalty0 (4):\penalty0 402, 2017.

\bibitem[Doersch(2021)]{doersch2021tutorial}
C.~Doersch.
\newblock Tutorial on variational autoencoders.
\newblock \emph{arXiv}, 1606.05908v3, 2021.
\newblock URL \url{https://arxiv.org/abs/1606.05908}.

\bibitem[Hennigan et~al.(2020)Hennigan, Cai, Norman, and
  Babuschkin]{haiku2020github}
T.~Hennigan, T.~Cai, T.~Norman, and I.~Babuschkin.
\newblock {H}aiku: {S}onnet for {JAX}, 2020.
\newblock URL \url{http://github.com/deepmind/dm-haiku}.

\bibitem[Hessel et~al.(2020)Hessel, Budden, Viola, Rosca, Sezener, and
  Hennigan]{optax2020github}
M.~Hessel, D.~Budden, F.~Viola, M.~Rosca, E.~Sezener, and T.~Hennigan.
\newblock Optax: composable gradient transformation and optimisation, in jax!,
  2020.
\newblock URL \url{http://github.com/deepmind/optax}.

\bibitem[Higgins et~al.(2017)Higgins, Matthey, Pal, Burgess, Glorot, Botvinick,
  Mohamed, and Lerchner]{higgins2016beta}
I.~Higgins, L.~Matthey, A.~Pal, C.~Burgess, X.~Glorot, M.~Botvinick,
  S.~Mohamed, and A.~Lerchner.
\newblock beta-vae: Learning basic visual concepts with a constrained
  variational framework.
\newblock In \emph{Proceedings of the International Conference on Learning
  Representations}, 2017.

\bibitem[Hinton et~al.(1995)Hinton, Dayan, Frey, and Neal]{hinton1995wake}
G.~E. Hinton, P.~Dayan, B.~J. Frey, and R.~M. Neal.
\newblock The" wake-sleep" algorithm for unsupervised neural networks.
\newblock \emph{Science}, 268\penalty0 (5214):\penalty0 1158--1161, 1995.

\bibitem[Hoffman and Johnson(2016)]{hoffman2016elbo}
M.~D. Hoffman and M.~J. Johnson.
\newblock Elbo surgery: yet another way to carve up the variational evidence
  lower bound.
\newblock In \emph{Workshop in Advances in Approximate Bayesian Inference,
  NeurIPS}, 2016.

\bibitem[Ioffe and Szegedy(2015)]{ioffe2015batch}
S.~Ioffe and C.~Szegedy.
\newblock Batch normalization: Accelerating deep network training by reducing
  internal covariate shift.
\newblock In \emph{International conference on machine learning}, pages
  448--456. PMLR, 2015.

\bibitem[Jumper et~al.(2020)Jumper, Evans, Pritzel, Green, Figurnov,
  Tunyasuvunakool, Ronneberger, Bates, Žídek, Bridgland, Meyer, Kohl,
  Potapenko, Ballard, Cowie, Romera-Paredes, Nikolov, Jain, Adler, Back,
  Petersen, Reiman, Steinegger, Pacholska, Silver, Vinyals, Senior,
  Kavukcuoglu, Kohli, and Hassabis]{jumper2020}
J.~Jumper, R.~Evans, A.~Pritzel, T.~Green, M.~Figurnov, K.~Tunyasuvunakool,
  O.~Ronneberger, R.~Bates, A.~Žídek, A.~Bridgland, C.~Meyer, S.~A.~A. Kohl,
  A.~Potapenko, A.~J. Ballard, A.~Cowie, B.~Romera-Paredes, S.~Nikolov,
  R.~Jain, J.~Adler, T.~Back, S.~Petersen, D.~Reiman, M.~Steinegger,
  M.~Pacholska, D.~Silver, O.~Vinyals, A.~W. Senior, K.~Kavukcuoglu, P.~Kohli,
  and D.~Hassabis.
\newblock High accuracy protein structure prediction using deep learning.
\newblock In \emph{Fourteenth Critical Assessment of Techniques for Protein
  Structure Prediction (Abstract Book)}, pages 22--24, 2020.
\newblock URL
  \url{https://predictioncenter.org/casp14/doc/CASP14_Abstracts.pdf}.

\bibitem[Kimanius et~al.(2021)Kimanius, Zickert, Nakane, Adler, Lunz,
  Sch{\"o}nlieb, {\"O}ktem, and Scheres]{kimanius2021exploiting}
D.~Kimanius, G.~Zickert, T.~Nakane, J.~Adler, S.~Lunz, C.-B. Sch{\"o}nlieb,
  O.~{\"O}ktem, and S.~H. Scheres.
\newblock Exploiting prior knowledge about biological macromolecules in cryo-em
  structure determination.
\newblock \emph{IUCrJ}, 8\penalty0 (1), 2021.

\bibitem[Kingma and Ba(2015)]{Kingma2015}
D.~P. Kingma and J.~Ba.
\newblock Adam: A method for stochastic optimization.
\newblock In \emph{Proceedings of the International Conference on Learning
  Representations}, 2015.

\bibitem[Kingma and Welling(2019)]{Kingma2019}
D.~P. Kingma and M.~Welling.
\newblock An introduction to variational autoencoders.
\newblock \emph{Foundations and Trends in Machine Learning}, 12\penalty0
  (4):\penalty0 307–392, 2019.
\newblock \doi{10.1561/2200000056}.

\bibitem[Lyumkis et~al.(2013)Lyumkis, Brilot, Theobald, and
  Grigorieff]{lyumkis2013likelihood}
D.~Lyumkis, A.~F. Brilot, D.~L. Theobald, and N.~Grigorieff.
\newblock Likelihood-based classification of cryo-em images using frealign.
\newblock \emph{Journal of structural biology}, 183\penalty0 (3):\penalty0
  377--388, 2013.

\bibitem[Nakane et~al.(2018)Nakane, Kimanius, Lindahl, and
  Scheres]{nakane2018characterisation}
T.~Nakane, D.~Kimanius, E.~Lindahl, and S.~H. Scheres.
\newblock Characterisation of molecular motions in cryo-em single-particle data
  by multi-body refinement in relion.
\newblock \emph{Elife}, 7:\penalty0 e36861, 2018.

\bibitem[Punjani and Fleet(2021)]{punjani20213d}
A.~Punjani and D.~J. Fleet.
\newblock 3d variability analysis: Resolving continuous flexibility and
  discrete heterogeneity from single particle cryo-em.
\newblock \emph{Journal of Structural Biology}, 213\penalty0 (2):\penalty0
  107702, 2021.

\bibitem[Punjani et~al.(2017)Punjani, Rubinstein, Fleet, and
  Brubaker]{punjani2017cryosparc}
A.~Punjani, J.~L. Rubinstein, D.~J. Fleet, and M.~A. Brubaker.
\newblock cryosparc: algorithms for rapid unsupervised cryo-em structure
  determination.
\newblock \emph{Nature methods}, 14\penalty0 (3):\penalty0 290--296, 2017.

\bibitem[Ruff et~al.(2018)Ruff, Muretta, Thompson, Lake, Cyphers, Albanese,
  Hanson, Behr, Thomas, Chodera, et~al.]{ruff2018dynamic}
E.~F. Ruff, J.~M. Muretta, A.~R. Thompson, E.~W. Lake, S.~Cyphers, S.~K.
  Albanese, S.~M. Hanson, J.~M. Behr, D.~D. Thomas, J.~D. Chodera, et~al.
\newblock A dynamic mechanism for allosteric activation of aurora kinase a by
  activation loop phosphorylation.
\newblock \emph{Elife}, 7:\penalty0 e32766, 2018.

\bibitem[Rullg{\aa}rd et~al.(2011)Rullg{\aa}rd, {\"O}fverstedt, Masich,
  Daneholt, and {\"O}ktem]{rullgaard2011simulation}
H.~Rullg{\aa}rd, L.-G. {\"O}fverstedt, S.~Masich, B.~Daneholt, and
  O.~{\"O}ktem.
\newblock Simulation of transmission electron microscope images of biological
  specimens.
\newblock \emph{Journal of microscopy}, 243\penalty0 (3):\penalty0 234--256,
  2011.

\bibitem[Scheres(2012)]{scheres2012relion}
S.~H. Scheres.
\newblock Relion: implementation of a bayesian approach to cryo-em structure
  determination.
\newblock \emph{Journal of structural biology}, 180\penalty0 (3):\penalty0
  519--530, 2012.

\bibitem[Senior et~al.(2019)Senior, Evans, Jumper, Kirkpatrick, Sifre, Green,
  Qin, Žídek, Nelson, Bridgland, Penedones, Petersen, Simonyan, Crossan,
  Kohli, Jones, Silver, Kavukcuoglu, and Hassabis]{senior2019protein}
A.~W. Senior, R.~Evans, J.~Jumper, J.~Kirkpatrick, L.~Sifre, T.~Green, C.~Qin,
  A.~Žídek, A.~W.~R. Nelson, A.~Bridgland, H.~Penedones, S.~Petersen,
  K.~Simonyan, S.~Crossan, P.~Kohli, D.~T. Jones, D.~Silver, K.~Kavukcuoglu,
  and D.~Hassabis.
\newblock Protein structure prediction using multiple deep neural networks in
  the 13th critical assessment of protein structure prediction (casp13).
\newblock \emph{Proteins: Structure, Function, and Bioinformatics}, 87\penalty0
  (12):\penalty0 1141--1148, 2019.

\bibitem[Villani(2009)]{villani2009optimal}
C.~Villani.
\newblock \emph{Optimal transport: old and new}.
\newblock Springer, 2009.
\newblock \doi{10.1007/978-3-540-71050-9}.

\bibitem[Vulovi{\'c} et~al.(2013)Vulovi{\'c}, Ravelli, van Vliet, Koster,
  Lazi{\'c}, L{\"u}cken, Rullg{\aa}rd, {\"O}ktem, and Rieger]{vulovic2013image}
M.~Vulovi{\'c}, R.~B. Ravelli, L.~J. van Vliet, A.~J. Koster, I.~Lazi{\'c},
  U.~L{\"u}cken, H.~Rullg{\aa}rd, O.~{\"O}ktem, and B.~Rieger.
\newblock Image formation modeling in cryo-electron microscopy.
\newblock \emph{Journal of structural biology}, 183\penalty0 (1):\penalty0
  19--32, 2013.

\bibitem[Wagner et~al.(2019)Wagner, Merino, Stabrin, Moriya, Antoni, Apelbaum,
  Hagel, Sitsel, Raisch, Prumbaum, et~al.]{wagner2019sphire}
T.~Wagner, F.~Merino, M.~Stabrin, T.~Moriya, C.~Antoni, A.~Apelbaum, P.~Hagel,
  O.~Sitsel, T.~Raisch, D.~Prumbaum, et~al.
\newblock Sphire-cryolo is a fast and accurate fully automated particle picker
  for cryo-em.
\newblock \emph{Communications biology}, 2\penalty0 (1):\penalty0 1--13, 2019.

\bibitem[Zhong et~al.(2021)Zhong, Bepler, Berger, and Davis]{zhong2021cryodrgn}
E.~D. Zhong, T.~Bepler, B.~Berger, and J.~H. Davis.
\newblock Cryodrgn: reconstruction of heterogeneous cryo-em structures using
  neural networks.
\newblock \emph{Nature Methods}, 18\penalty0 (2):\penalty0 176--185, 2021.

\end{thebibliography}

\end{document}